\documentclass[12pt]{iopart}
\usepackage{iopams}
\usepackage{latexsym,epsf,cite}
\usepackage{epsfig,epic}
\usepackage{wrapfig}
\usepackage{multirow}
\newcommand{\eq}{\begin{equation}} 
\newcommand{\en}{\end{equation}} 
\newcommand{\eqa}{\begin{eqnarray}} 
\newcommand{\ena}{\end{eqnarray}}

\newcommand{\um}{\frac12}

\newcommand{\lan}{\langle} 
\newcommand{\ran}{\rangle}

\newcommand{\dep}{\partial}
\begin{document} 
\title{Finite temperature results on the 2$d$ Ising model with mixed perturbation}
\author{P.~Grinza\dag\footnote[3]{grinza@sissa.it} and A.~Rago\ddag\footnote[7]{rago@to.infn.it}}
\address{\dag\ International School for Advanced Studies (SISSA)
and INFN sezione di Trieste, 
via Beirut 2-4, 34014 Trieste, Italy}
\address{\ddag\ Dipartimento di Fisica  Teorica dell'Universit\`a di Torino
  and INFN sezione di Torino, 
via P.Giuria 1, I-10125 Torino, Italy}
\begin{abstract}
A numerical study of finite temperature features of thermodynamical observables is performed for the lattice 2$d$ Ising model. Our results support the conjecture that the Finite Size Scaling analysis employed in the study of integrable perturbation of Conformal Field Theory is still valid in the present case, where a non-integrable perturbation is considered.
\end{abstract} 
\maketitle
\section{Introduction} 
Quantum field theories (QFT) in two space-time dimensions at finite temperature are an appealing and complex subject which has attracted much attention from both theoretical and experimental point of view. It could be useful to recall, that a finite temperature QFT can be obtained by considering the model on a cylinder with the radius of compactification proportional to the inverse of the physical temperature. 

Since the 2$d$ Ising model fuses together the simplicity and the full features of more complex models, we considered the most general, relevant perturbation of such model (both thermal and magnetic) in order to cope with the problem of finite temperature corrections to thermodynamic observables (in particular, as it will be clear later, such corrections are nothing but the finite size scaling behaviour of the model as a function of the radius of compactification).

In this respect we recall that, taking advantage of integrability, the finite temperature study of the pure magnetic perturbation of the model was performed in \cite{Delfino:2001sz} where an expression for the one point functions was proposed in the framework of Form Factors. Such results were found to be in agreement with the high precision numerical analysis of \cite{Caselle:2002cs}.

In this paper, we investigate the finite temperature features of the model in the region of the coupling constants ($h$ representing the coupling to the spin operator, and $t=|\beta-\beta_c|/\beta$ the deviation from the critical temperature) where the thermal perturbation is small with respect to magnetic one, $t/h^{8/15} \to 0$.

The main idea underlying our approach is to evaluate the Finite Size Scaling (FSS) on the cylindric geometry of the main thermodynamic observables (magnetization, free energy, magnetic susceptibility) by means of high precision numerical data coming from numerical diagonalization of the Transfer Matrix. Consequently, the finite temperature corrections to such observables, i.e. the FSS behaviour, are given by standard Thermodynamic Bethe Ansatz (TBA) prescriptions.

Such an approach is known to work when the perturbations are separately switched on, giving rise to Integrable QFTs (see \cite{Delfino:2001sz,Caselle:2002cs,Caselle:2000nn}). The contemporary presence of both of them destroys the integrability, however since we are in the region of the phase diagram where $t/h^{8/15} \to 0$, we postulate that the functional forms coming from TBA still hold.

We showed that this expectation is indeed correct (at least within the precision of our data), and as a side result we were able to evaluate the infinite volume quantities with higher precision than before \cite{Grinza:2002pi}. It is important to point out that such functional form can be rigorously derived only when integrable theories are concerned, however we expect that when integrability is lost, they still hold under pair creation threshold which is just the regime explored in our simulations.

The work is organized as follows: in sect.~2 we briefly present the model and the numerical transfer matrix technique; sect.~3 is devoted to the study of the FSS of some observables of the Ising model in magnetic field; in sect.~4 we present our numerical results; finally, in sect.~5 we give our conclusion.   

\section{The model and the technique}
The Ising model in a magnetic field at an arbitrary temperature is defined by the partition function
\eqa
\label{defz}
Z(\beta, h_\ell)~=~\sum_{\sigma_i=\pm 1} e^{\beta \sum_{\langle n,m \rangle}\sigma_n \sigma_m + h_\ell 
\sum_n \sigma_n },
\ena 
where the spin variable $\sigma_n$ takes the values $\pm 1$; the notation
$\langle n,m \rangle$ represents nearest neighbor sites on the lattice; the
sites are labeled by $n = (n_o,n_1)$ and the two sizes of the square lattice
are $L$ and $R$ (they are taken to be different because our transfer matrix
calculations will treat the two directions asymmetrically); the total
number of sites of the lattice will be denoted as $N = L  R$. \\
The coupling $\beta$ is the inverse of the temperature, while the magnetic
perturbation is introduced by the coupling $h_\ell\equiv H \beta$, where $H$
is the magnetic coupling. This model undergoes a second order phase
transition when $h_\ell=0$ and $\beta$ reaches its critical value $\beta_c$ \cite{Onsager:1943jn}
\eqa \beta_c\equiv\um \log (\sqrt 2 +1) = 0.4406868 \dots \,.\ena
The observables which we shall consider are the following: \\
the {\it Magnetization}
\eqa
M(\beta, h_\ell) = \frac1N \frac{\dep}{\dep h_\ell}(\log Z(\beta, h_\ell) )=
\frac1N \lan \sum_i \sigma_i \ran.
\ena
the {\it Magnetic Susceptibility} 
\eqa
\chi (\beta, h_\ell) = \frac{\dep M(\beta, h_\ell)}{\dep h_\ell}.
\ena
the {\it Free Energy}
\eqa
f(\beta, h_\ell) = \frac1N \log Z(\beta, h_\ell)-f_b.
\ena
where the bulk constant $f_b= 0.9296953982 \dots $ can be found in \cite{Onsager:1943jn}.
\vskip0.3cm
Since we will often refer to exact results coming from the QFT which describes the Ising model in the scaling limit, we recall that its action is given by
\eqa
\label{IFT}
{\mathcal A_{\textrm{\tiny IM}}} = {\mathcal A_{\textrm{\tiny CFT}}}  + \tau \int d^2 x \epsilon(x) + h \int d^2 x \sigma(x) 
\ena
where the first term is the action of the Minimal Unitary CFT describing the critical point of the model \cite{Belavin:1984vu}, $h$ is the coupling constant
associated to the magnetic field and $\tau$ gives the deviation from the critical temperature (for a review on this subject see \cite{Delfino:2003yr}). Such a theory is integrable either when $h=0$, $\tau \neq 0$ and one can prove that it describes a massive Majorana fermion, or in the case $h \neq 0$, $\tau = 0$ which turns out to be a non-trivial theory of eight self-conjugated particles whose spectrum follows the numerology of the exceptional group $E_8$ as proved by Al.\ B.\ Zamolodchikov in \cite{Zamolodchikov:1989fp}. Simple arguments from CFTs are sufficient to conclude that in the general case $h \neq 0$, $\tau \neq 0$ integrability\footnote{We would like to stress that we will always refer to the integrability of the continuum theory because we will use exact results coming from the latter. We are not at all interested in the integrability of the lattice model since our approach is numeric and hence it is not related to the lattice integrability of (\ref{defz}).} is lost \cite{Mussardo:uc}.
\vskip0.4cm
In order to analyze the finite temperature behaviour of the model our choice was to employ the numerical
transfer matrix technique. The basic idea is to rewrite the Boltzmann weight by means of the so-called transfer matrix $T$, which turns out to be a a positive symmetric $2^R \times 2^R$ matrix such that   
\eqa
\label{prtfnt}
Z(\beta, h_\ell) = \sum_{\sigma_i=\pm 1} e^{\beta \sum_{\langle n,m \rangle}\sigma_n \sigma_m + h_\ell } = \textrm{tr}\;T^{L}~=~\sum_i \lambda_i^{L}
\ena
where the $\lambda_i$ are the (real) eigenvalues of $T$ (details can be found in \cite{kwan,baxterbook,tramat}). 
The numerical computation of eigenvalues and eigenvectors of the
transfer matrix enables us to compute all the observables
we need, provided that we specify the values of $h_\ell$, $\beta$ and $R$. 
As an example, the free energy
\eqa
f(\beta, h_\ell) & =& \frac{1}{R \; L} \log Z(\beta, h_\ell)
\ena
can be written, in the limit $L \to \infty$, as
\eqa
f(\beta, h_\ell) & \sim &  \frac{1}{R} \log \lambda_{max}
\ena
which is exactly what we need for the numerical study of the thermodynamic observable as functions of the length of the cylinder $R$.
\section{Finite temperature results, the integrable case}
Taking advantage of the integrability of the model (\ref{IFT}) when ($\tau=0, h \neq0$), Delfino \cite{Delfino:2001sz} applied Form Factor method to propose a  compact expression for the finite size (i.e. finite temperature) corrections of the one point functions of the Ising model with pure magnetic perturbation (see also \cite{llss96,m2001,lm99} for a discussion on the finite temperature analysis of 2$d$ integrable Quantum Field Theory).\\
Using these results, a given one point function $\lan \Phi \ran_R$ calculated on a cylinder of radius $R$ is related to its infinite plane value $\lan \Phi \ran_\infty$ by
\eqa
\label{ft}
\frac{\lan \Phi \ran_R}{\lan \Phi \ran_\infty} \ = \
1+ \frac{1}{\pi}\; \sum^{3}_{i=1} \mathcal{A}_i^{\Phi} \; K_0 (r_i) +  \mathcal{O}(e^{-2 r_1 }) 
\ena
where $r_i=m_i R$ and $m_i$, with $i=1,2,3$, are the first three masses of Zamoldchikov's mass spectrum \cite{Zamolodchikov:1989fp}. \\
The main result of \cite{Delfino:2001sz} is to compute exactly the universal constants $A_i^{\Phi}$.
A numerical check of (\ref{ft}) has been performed by Caselle et al. \cite{Caselle:2002cs} using Transfer Matrix technique to study the Ising model perturbed with a magnetic field.
As already pointed out in \cite{Delfino:2001sz}, the one point function of the perturbing operator $\sigma$ can be obtained by TBA calculations. 
In fact, following TBA prediction for the FSS behavior of the free energy, it is possible to write down
\eqa
f(R) \ = \ f(\infty) + \sum^{3}_{i=1}\frac{E (r_i)}{R} + \mathcal{O}(e^{-2 r_1})
\ena
where $E (r_i)$ is given by
\eqa
E (r_i) = - \frac{m_i}{\pi} \;  K_1 (r_i).
\ena
and $K_1 (r)$ is the modified Bessel function of the second kind.\\ 
Hence, for the free energy we have
\eqa
\frac{f(R)}{f(\infty)} = 1 -  \sum^{3}_{i=1} 
\frac{\mathcal{A}_i^{f}}{\pi} 
\; \frac{ K_1 (r_i)}{r_i}+  \mathcal{O}(e^{-2 r_1});
\hspace{2cm}
\mathcal{A}_i^{f}=
\frac{{\mathcal C}_i^2}{A_f }
\ena
which gives us the first few orders (under the lowest particle pair creation threshold) of the finite size corrections;
the constants ${\mathcal C}_i$ are given by Zamolodchikov's mass spectrum (see ref.\ \cite{Zamolodchikov:1989fp}).
A first derivative with respect to the magnetic field gives us just 
expression (\ref{ft}) for the magnetization \footnote{The constants ${\mathcal A}_i^\sigma$ coincide with the coefficients $A_i^\sigma$ found by Delfino with the Form Factor approach in \cite{Delfino:2001sz}.}
\eqa
\frac{\lan  \sigma \ran_R}{\lan \sigma \ran_\infty} \ = \
1+ \sum^{3}_{i=1}  \frac{\mathcal{A}_i^{\sigma}}{\pi} 
 \; K_0 (r_i) + \mathcal{O}(e^{-2 r_1});
\hspace{2cm}
\mathcal{A}_i^{\sigma}=
\frac{{\mathcal C}^2_i}{2 A_f}
\ena
then a further derivative gives us the functional form of the finite size corrections for the magnetic susceptibility      
\eq
\frac{\chi(R)}{\chi(\infty)} = 1 -\sum^{3}_{i=1} \frac{\mathcal{A}_i^{\chi}}{\pi}
 \; [ K_0 (r_i)  -  8\; r_i\; K_1 (r_i) ] +  \mathcal{O}(e^{-2 r_1});
\hspace{0.5cm}
\mathcal{A}_i^{\chi}=
\frac{{\mathcal C}^2_i}{2 A_f}.
\en
The exact values of the constants are given in table \ref{par_val}.
\begin{table}[ht]
\begin{center}
\begin{tabular}{c}
\hline
\hline
\\[-3mm]
$\mathcal{C}_1=4.40490858\dots$\\[1mm]
$\mathcal{C}_2=7.12729179\dots$\\[1mm]
$\mathcal{C}_3=8.76155605\dots$\\[1mm]
$A_f=1.19773338\dots$\\[1mm]
\hline
\hline
\end{tabular}
\end{center}
\caption{Numerical values of the constants.}
\label{par_val}
\end{table}\\
The aim of the next section is twofold: on the one hand, we perform a numerical check for the finite size formulas for free energy and magnetic susceptibility in the case of the integrable model (for internal energy and magnetization this was proved in \cite{Caselle:2002cs}); on the other hand we check
that all the functional forms also hold in the case of the mixed perturbation considered in \cite{Grinza:2002pi}. The latter case implies that 
the constants acquire a non-trivial dependence with respect to both $h_\ell$
and $t$. Their functional form can be derived by the same scaling arguments 
employed in \cite{Grinza:2002pi}.
\section{Numerical analysis}
In order to check the behaviour of our data coming from the numerical diagonalization of the transfer matrix, we use the effective procedure presented in our previous work \cite{Grinza:2002pi} and in \cite{Caselle:1999bx}.

First, for each value of $h_\ell$ we fit the value of $\lan \Phi \ran_R$ according to
the functional form of the finite size corrections found in the previous section. As an example we can consider the susceptibility, hence the fitting function is given by (see previous section)
\eq
\frac{\chi(R)}{\chi(\infty)} = 1 -\sum^{3}_{i=1} \frac{\mathcal{A}_i^{\chi}}{\pi}
 \; [ K_0 (r_i)  -  8\; r_i\; K_1 (r_i) ] +  \mathcal{O}(e^{-2 r_1});
\en
where we keep the quantities $\mathcal{A}_i^{\chi}$ and $\chi(\infty)$ as free parameters. Our fitting procedure follows a very stringent list of acceptance criteria as presented in \cite{Grinza:2002pi}. The first non trivial result is that the fits are acceptable (for all the observable considered here) even if the temperature is not set to its critical value. In this case the behaviour of the observables as function of $R$ cannot be proved to hold rigorously because the corresponding QFT is not integrable. Notwithstanding this, in the regime $t/h^{8/15} \to 0$ considered here and within our numerical precision, such functional forms continue to hold. 

To the best of our knowledge the present is the first evidence about the validity of formulas like that of previous section for non-integrable theories. It could suggest that, when small perturbations of integrable QFTs are considered, then the finite temperature behaviour of the thermodynamic observables remains the same and the only price to pay is a change in the constants ${\mathcal A}_i^\phi$ which acquire a non-trivial dependence with respect to the couplings $t$ and $h_\ell$.   
The following quantitative results were found: 
\begin{enumerate}
\item{The numerical results for the infinite $R$ observables $f(\infty)$, $\langle \sigma \rangle_\infty$, $\chi(\infty)$ are not only in complete agreement with respect to the estimates of \cite{Grinza:2002pi}, but they show an improvement in their precision of about one order of magnitude. As an example of this we quote the following results for the susceptibility   
\vskip0.3cm
\underline{$\beta=\beta_c$, $h=0.07$}
\eqa
\chi (\beta_c,0.07)=0.7574405(3)
\ena
compared with the value obtained in \cite{Grinza:2002pi}
\eqa
\chi (\beta_c,0.07)=0.757441(1);
\ena
\underline{$\beta=0.4349\dots$, $h=0.07$}
\eqa
\chi (0.07,0.4349\dots)=0.8465333(3)
\ena
compared with the value obtained in \cite{Grinza:2002pi}
\eqa
\chi (0.07,0.4349\dots)=0.846533(1).
\ena
We stress that it is remarkable that such an improvement comes independently of the fact that $\beta$ is fixed at its critical value or not. Moreover such an improvement in precision could be expected since in this latter case we used the functional form of the finite temperature corrections instead of fitting with a set of generic exponentially decaying functions.}
\item{As anticipated, we supposed that the constants ${\mathcal A}_i^\phi$ are functions of $t$ and $h_\ell$. In order to test such a conjecture, we first define the following subtracted quantities in order to discard all the corrections which depends only on the magnetic field $h_\ell$,
\begin{equation}
\mathcal{A}^\phi_i (t) |_{h_\ell} - \mathcal{A}^\phi_i (0) |_{h_\ell} =\mathcal{J}^\phi_{1,i}({h_\ell})~ t +  \mathcal{J}^\phi_{2,i}({h_\ell})~ t^2 +  \mathcal{J}^\phi_{3,i}({h_\ell})~t^3 + \dots
\label{sott}
\end{equation}
As pointed out in \cite{Grinza:2002pi}, it is a crucial step in order to compute the leading corrections due to the thermal perturbation. In this way we are able to fit our data with (\ref{sott}). The precision of our data enable us to compute only the linear term of such an expansion.

By means of the expansion for the scaling functions we are in the position to fit the amplitude $\mathcal{J}^\mathcal{O}_1({h_\ell})$ as a function of the magnetic field alone. We recall that, in principle, the functional form of the $\mathcal{J}^\phi_{1,i}({h_\ell})$ is non-trivial to be obtained, however, they can be determined by the same Renormalization Group and CFT arguments given in \cite{Caselle:2002cs,Grinza:2002pi} in order to construct the scaling functions of the thermodynamic observables. The results are 
\begin{itemize}
\item[]{\underline{Free Energy}: The fit with the scaling function
\eqa
\mathcal{J}^f_{1,1}({h_\ell}) = \frac{\mathcal{W}_1}{h_\ell^\frac{8}{15}}+
\mathcal{W}_2~h_\ell^\frac{2}{15}+
\mathcal{W}_3~h_\ell^\frac{8}{15}+\dots
\ena
and we can give a reliable estimate only of the leading term of the expansion
\eqa
\mathcal{W}_1=90(8).
\ena
}
\item[]{\underline{Magnetization}: The fit with the scaling function 
\eqa
\mathcal{J}^\sigma_{1,1}({h_\ell}) = \frac{\mathcal{R}_1}{h_\ell^\frac{8}{15}}+
\mathcal{R}_2~h_\ell^\frac{2}{15}+
\mathcal{R}_3~h_\ell^\frac{8}{15}+\dots
\ena
gives us
\eqa
2.09<\mathcal{R}_1<2.15.
\ena
}
\end{itemize}  
The other observables were outside our numerical precision. 
}
\item{Finally, a further check about the validity is given by the analysis of the data at the critical temperature $\beta_c$. In this case the scaling function for the coefficients ${\mathcal A}_1^\phi$ depend only on the magnetic field $h_\ell$ and is given by
\eqa
\hat{\mathcal{A}}_1^{\Phi}(h_\ell)=\mathcal{A}_1^{\Phi}(1+
b_{1,\Phi}|h_\ell|^{\frac{16}{15}}+
b_{2,\Phi}|h_\ell|^{\frac{22}{15}}) +\dots .
\ena
Employing them as fitting functions, we find
\eqa
8.07<&\mathcal{A}_1^{\chi}&<8.20\\
16.12<&\mathcal{A}_1^{f}&<16.28\hspace{1cm}  
\ena
where the numerical values of the amplitudes perfectly agree with their theoretical
counterparts: $\mathcal{A}_1^{\chi}=8.09997\dots$ and $\mathcal{A}_1^{f}=16.1999\dots$ computed according to the formul\ae~given in the previous section. Such results can be considered as further checks (beside the results of ref. \cite{Caselle:2002cs} where magnetization and internal energy were considered) of the results coming from form factors. 
}
\end{enumerate}

The complete list of the numerical values of the parameters we used for the numerical diagonalization of the Transfer Matrix is given in table \ref{tb1}. A significant sample of the data employed in the computation of finite temperature corrections is given in table \ref{tb2}.
\begin{table}
\begin{center}
\begin{tabular}{l}
\hline
\hline
\multicolumn{1}{c}{$t$}\\
\hline
0.0130481\\0.0119726\\0.0108971\\0.0087367\\0.0076518\\0.0065668\\0.0043875\\0.0021985\\0.0010992\\0.0\\
\hline
\hline
\end{tabular}
\hspace{5mm}
\begin{tabular}{l}
\hline
\hline
\multicolumn{1}{c}{$h_\ell$}\\
\hline
0.01\\
0.02\\
0.03\\
0.04\\[1.5mm]
$\vdots$\\[1.5mm]
0.17\\
0.18\\
0.19\\
0.20\\
\hline
\hline
\end{tabular}
\hspace{5mm}
\begin{tabular}{l}
\hline
\hline
\multicolumn{1}{c}{$R$}\\
\hline
9\\
10\\
11\\
12\\[1.5mm]
$\vdots$\\[1.5mm]
17\\
18\\
19\\
20\\
\hline
\hline
\end{tabular}
\end{center}
\caption{Value of the parameters used for the diagonalization of the Transfer Matrix.}
\label{tb1}
\end{table}

\begin{table}
\begin{center}
\begin{tabular}{l}
\hline
\hline
\multicolumn{1}{c}{$R$} \\
\hline
9  \\
10 \\
11\\
12 \\
13  \\
14 \\
15 \\
16 \\
17  \\
18  \\
19 \\
20  \\
\hline
\hline
\end{tabular}
\hspace{5mm}
\begin{tabular}{l}
\hline
\hline
\multicolumn{1}{c}{$\chi$}\\
\hline
0.860965432368 \\
0.852426578479 \\
0.848936912437\\
0.847512060489\\
0.846931329088\\
0.846695116084\\
0.846599363455\\
0.846560454648\\
0.846544860407\\
0.846538199957\\
0.846535213111\\
0.846532714557 \\
\hline
\hline
\end{tabular}
\end{center}
\caption{Magnetic susceptibility as function of the compactification radius $L$ ($h_\ell=0.07,\beta=0.4349\dots$).}
\label{tb2}
\end{table}

\section{Conclusions}
\label{sec:discandconcl}
Let us recall our main results. Our numerical study confirms the conjecture that the functional forms of the finite temperature corrections derived by means of Thermodynamic Bethe Ansatz for the Ising model with magnetic perturbation at $t=0$ ($T=T_c$) still holds when $t \neq 0$ supposing that $t/h^{8/15} \to 0$. 
We stress that, the functional form of finite temperature corrections works well even if integrability is lost because we explored a region of the phase space where anelastic effect can be neglected. As explained in \cite{Delfino:1996xp} the thermal perturbation results only in a correction to the mass spectrum of the integrable Ising field theory in magnetic field; no other effects due to the loss of integrability are expected. 

Finally, it could be worth to notice that, the high precision of our data allowed us to achieve the following numerical results: first we obtained the infinite volume estimates of the observables with typically one order of magnitude of precision more than the method previously employed in \cite{Grinza:2002pi}; second we get information about the first subleading correction due to the reduced temperature. 

\end{document}